\documentclass[12pt]{article}
\usepackage[centertags]{amsmath}
\usepackage{amssymb,amsfonts}
\usepackage[comma,super]{natbib}
\usepackage{latexsym}
\begin{document}
\begin{center}
{\Large \bf Radiating relativistic matter in geodesic motion} \\
\vspace{1.5cm} {\bf S. Thirukkanesh\footnote{Permanent address:
Department of Mathematics, Eastern
University, Chenkalady, Sri Lanka.} and S. D. Maharaj\footnote{Electronic mail: maharaj@ukzn.ac.za}}\\
Astrophysics and Cosmology Research Unit,\\
School of Mathematical Sciences,\\
University of KwaZulu-Natal,\\
Private Bag X54001,\\
Durban 4000,\\
South Africa.\\
\vspace{1.5cm} {\bf Abstract}\\
\end{center}
We study the gravitational behaviour of a spherically symmetric
radiating star when the fluid particles are in geodesic motion. We
transform the governing equation  into a simpler form which allows
for a general analytic treatment. We find that Bernoulli, Riccati
and confluent hypergeometric equations are possible. These admit
solutions in terms of elementary functions and special functions.
Particular models contain the Minkowski spacetime and the Friedmann
dust spacetime as limiting cases. Our infinite family of solutions
contains specific models found previously. For a particular metric
we briefly investigate the physical features, derive the temperature
profiles and plot the behaviour of the casual and acasual
temperatures.

\section{Introduction}
Relativistic models of radiating stars are useful in the
investigation of cosmic censorship hypothesis, gravitational
collapse with dissipation, formation of superdense matter, dynamical
stability of radiating matter and temperature profiles in the
context of irreversible thermodynamics. The general model,
incorporating all necessary physical requirements and variables, is
complicated and difficult to solve; the treatments of Herrera
\emph{et al} [1] and Di Prisco \emph{et al} [2] involving physically
meaningful charged spherically symmetric collapse with shear and
dissipation illustrate the complexity of the processes. To solve the
field equations, and to find tractable forms for the gravitational
and matter variables, we need to make simplifying assumptions. De
Oliviera \emph{et al} [3] proposed a radiating model in which an
initial static configuration leads to collapse. This approach may be
adapted to describe the end state of collapse as shown by Govender
\emph{et al} [4].  In a recent treatment Herrera \emph{et al} [5]
proposed a model in which the form of Weyl tensor was highlighted
when studying radiative collapse with an approximate solution.
Maharaj and Govender [6], Herrera \emph{et al} [7] and Misthry
\emph{et al} [8] showed that it is possible to solve the field
equations and boundary conditions exactly in this scenario. For
recent treatments involving collapse with equations of state and
formation of black holes see Goswami and Joshi [9], [10].

A useful approach in understanding the effects of dissipation is due
to Kolassis \emph{et al} [11] in which the fluid trajectories are
assumed to be geodesic. In the limit, in the absence of heat flow,
the interior Friedmann dust solution was regained. This solution
formed the basis for many investigations involving the physical
behaviour such as the rate of collapse, surface luminosity and
temperature profiles. These include the analytic model of radiating
spherical gravitational collapse with neutrino flux by Grammenos and
Kolassis [12], the model describing realistic astrophysical
processes with heat flow by Tomimura and Nunes [13], and models
undergoing collapse with heat flow as a possible mechanism for
gamma-ray bursts by Zhe  \emph{et al} [14]. Herrera \emph{et al}
[15] considered geodesic fluid spheres in coordinates which are not
comoving but with anisotropic pressures. Govender \emph{et al} [16]
showed that the behaviour of the temperature in casual
thermodynamics for geodesic motion produces higher central
temperatures than the Eckart theory.  The first exact solution with
shear, satisfying the boundary conditions,  was obtained by Naidu
\emph{et al} [17]  by considering geodesic fluid trajectories. Later
Rajah and Maharaj [18] extended this treatment and obtained classes
of models which are nonsingular at the centre.

It is clear that the assumption of geodesic motion is physically
acceptable and has been used by other investigators in attempts to
describe realistic astrophysical processes. In this paper we attempt
to perform a systematic treatment on the governing equation at the
boundary for shear-free collapse by assuming the geodesic motion of
the fluid particle. Our intention is to show that the nonlinear
boundary condition may be analysed systematically to produce an
infinite family of exact solutions. In Section 2, we present the
model governing the description of a radiating star using the
Einstein field equations together with the junction conditions. We
show that it is possible to  transform the junction condition to a
Bernoulli equation and a  Riccati equation. Solutions are obtained
in terms of elementary functions in Section 3. In Section 4, we show
that the boundary condition, under relevant assumptions, can be
written in the form of a confluent hypergeometric equation. We
demonstrate that an infinite family of solutions in terms of
elementary functions are possible. In Section 5, we obtain the
explicit form for the causal temperature using the truncated form of
the Maxwell-Cattaneo heat transport equation for a particular
metric. This illustrates that the simple forms for the gravitational
potentials obtained in this paper are physically plausible. Some
concluding statements are made in Section 6.

\section{The model}
We analyse a spherically symmetric relativistic radiating star
undergoing shear-free gravitational collapse. This assumption is
reasonable when modelling a radiating star in relativistic
astrophysics. If we suppose that the particle trajectories are
geodesic then the acceleration vanishes. Then the line element, for
the matter distribution interior to the boundary of the radiating
star, is given by
\begin{equation}
\label{eq:e1} ds^2 = - dt^2 + B^2 \left[ dr^2 + r^2 (d\theta^2 +
\sin^2\theta d\phi^2)\right]
\end{equation}
where $B=B(r,t)$ is the only surviving metric function. The energy
momentum tensor including radiation for the interior spacetime is
given by
\begin{equation}
\label{eq:e2} T_{ab}= (\rho + p)u_a u_b +p g_{ab} +q_a u_b +q_b
u_a
\end{equation}
where the energy density $\rho$, the pressure $p$ and the heat flow
vector $\mathbf{q}$ are measured relative to the timelike fluid
4-velocity $u^a = \delta^a_0$. The heat flow vector takes the form
$q^a =(0,q,0,0)$ since $\mathbf{q}\cdot \mathbf{u} =0$ for heat flow
which  is radially directed.

The nonzero components of Einstein field equations, for the line
element (\ref{eq:e1}) and the energy momentum tensor (\ref{eq:e2}),
can be written as
\begin{subequations}
\label{eq:e3}
\begin{eqnarray}
\label{eq:e3a} \rho &=& 3 \frac{\dot{B}^2}{B^2} - \frac{1}{B^2}
\left(2 \frac{B''}{B} - \frac{{B'}^2}{B^2} +
\frac{4}{r}\frac{B'}{B} \right),\\
\label{eq:e3b} p &=& - 2 \frac{\ddot{B}}{B} -
\frac{\dot{B}^2}{B^2} + \frac{1}{B^2} \left(\frac{{B'}^2}{B^2}
+\frac{2}{r}\frac{B'}{B} \right),\\
\label{eq:e3c} p &=& -2 \frac{\ddot{B}}{B}-\frac{\dot{B}^2}{B^2}
+\frac{1}{B^2} \left(\frac{B''}{B}
 - \frac{{B'}^2}{B^2}+\frac{1}{r}\frac{B'}{B}\right),\\
\label{eq:e3d} q & = & - \frac{2}{B^2}\left( - \frac{\dot{B}'}{B} +
\frac{B'\dot{B}}{B^2} \right),
\end{eqnarray}
\end{subequations}
where dots and primes denote differentiation with respect to time
$t$ and $r$ respectively. Equating (\ref{eq:e3b}) and (\ref{eq:e3c})
we obtain the condition
\begin{equation}
\label{eq:e4}
\left(\frac{1}{B}\right)''=\frac{1}{r}\left(\frac{1}{B}\right)'
\end{equation}
which is the  condition of pressure isotropy. Equation (\ref{eq:e4})
is integrable and we obtain
\begin{equation}
\label{eq:e5} B= \frac{d}{C_2 (t) - C_1 (t)r^2 }
\end{equation}
where $C_1 (t)$ and $C_2 (t)$ are functions of time, and $d$ is a
constant. As the functional form for the potential $B$ is specified
the matter variables $\rho, p$ and $q$ are known quantities, and the
system (3) has been solved in principle.

The interior spacetime (\ref{eq:e1}) has to be matched across the
boundary $r=b$ to the exterior Vaidya spacetime
\begin{equation}
\label{eq:e6} ds^2 = - \left(1 - \frac{2 m(v)}{R}\right)dv^2 -2 dv
dR + R^2 (d\theta^2 +\sin^2\theta d\phi),
\end{equation}
where $m(v)$ denotes the mass of the star as measured by an observer
at infinity. The hypersurface at the boundary is denoted by
$\Sigma$. The matching of the line elements (\ref{eq:e1}) and
(\ref{eq:e6}), and matching of the extrinsic curvature at the
surface of the star, leads to a set of equations. The boundary
conditions at $\Sigma$ have the form
\begin{subequations}
\label{eq:e7}
\begin{eqnarray}
dt & = & \left[ \left( 1 - \frac{2m}{R} +2
\frac{dR}{dv}\right)^{1/2} dv \right]_{\Sigma}, \\
(r B)_{\Sigma} &=& R_{\Sigma}, \\
\label{eq:e7c} p_{\Sigma} &=& (q B)_{\Sigma}, \\
\label{eq:e7d}\left[ m(v)\right]_{\Sigma} &= & \left[ \frac{r^3}{2}
\left( \dot{B}^2 B - \frac{{B'}^2}{B}\right)- r^2
B'\right]_{\Sigma},
\end{eqnarray}
\end{subequations}
where the subscript means that the relevant quantities are evaluated
on $\Sigma$.

From (\ref{eq:e3}), (\ref{eq:e5}) and (\ref{eq:e7c}) we generate the
condition
\begin{eqnarray}
 -4 db(\dot{C_1}C_2 -C_1 \dot{C_2})(C_1 b^2 -C_2)-
4 C_1 C_2(C_1 b^2 -C_2)^2 && \nonumber\\
\label{eq:e8} -2d^2 (\ddot{C_1} b^2 -\ddot{C_2})(C_1 b^2
-C_2)+5d^2(\dot{C_1}b^2-\dot{C_2})^2&=&0.
\end{eqnarray}
Effectively (\ref{eq:e8}) results from the nonvanishing of the
pressure gradient across the hypersurface $\Sigma$.  Equation
(\ref{eq:e8}) governs the dynamical evolution of shear-free
radiating stars in which fluid trajectories are geodesic. To
complete the description in this particular radiating model we need
to explicitly solve the differential equation (\ref{eq:e8}).

\section{Generating Analytic Solutions}
A particular solution to (\ref{eq:e8}) was found by Kolassis
\emph{et al} (1988) by inspection. We show that it is possible to
transform (\ref{eq:e8}) into familiar differential equations which
admit solutions in closed form. Our method is a more systematic
approach in solving equation (\ref{eq:e8}). In this approach we let
\begin{equation}
\label{eq:e9} C_1 b^2 - C_2 =u(t).
\end{equation}
On substituting (\ref{eq:e9}) into (\ref{eq:e8}) we can write
\begin{equation}
\label{eq:e10} 4 bdu^2\dot{C_1}+4(u^2-bd\dot{u})uC_1
-4b^2u^2C_1^2=d^2(2u\ddot{u}-5{\dot{u}}^2).
\end{equation}
Equation (10) is simpler than (\ref{eq:e8}) and can be viewed as a
first order differential equation in the variable $C_1$. In general,
(\ref{eq:e10}) is a Riccati equation (in $C_1$), and is difficult
solve in the above form without simplifying assumptions. For the
integration  of (\ref{eq:e10}), in terms of elementary functions, we
consider the following two cases:

\subsection{Bernoulli equation}
We set
\begin{equation}
\label{eq:e11} 2u\ddot{u}-5{\dot{u}}^2=0
\end{equation}
so that the function $u$ is given by
\begin{equation}
\label{eq:e12}u=\alpha~~\mbox{or}~~u=\beta (t+\gamma)^{-2/3},
\end{equation}
where $\alpha, \beta$ and $\gamma$ are real constants. With the
assumption (\ref{eq:e11}), (\ref{eq:e10}) becomes
\begin{equation}
\label{eq:e13} 4 bdu^2\dot{C_1}+4(u^2-bd\dot{u})uC_1
-4b^2u^2C_1^2=0.
\end{equation}
Equation (\ref{eq:e13}) is nonlinear but is a Bernoulli equation
which can be linearised in general.

When $u=\alpha$,  equation (\ref{eq:e13}) becomes
\begin{equation}
\label{eq:e14} \dot{C_1} +\frac{\alpha}{bd}C_1 -\frac{b}{d}C_1^2=0
\end{equation}
which is a Bernoulli equation with constant coefficients. The
solution of (\ref{eq:e14}) is given by
\[C_1=\frac{\alpha}{b^2-\exp\left(\frac{\alpha (t+e)}{bd}\right)},
\]
where $e$ is the constant of integration. Consequently the remaining
function $C_2$ has the form \[ C_2= \frac{\alpha
\exp\left(\frac{\alpha (t+e)}{bd}\right)}{b^2-\exp\left(\frac{\alpha
(t+e)}{bd}\right)}.
\]
Hence the interior line element (\ref{eq:e1}) has the specific form
\begin{equation}
\label{eq:e15} ds^2 = - dt^2 + \frac{d^2}{{\alpha}^2}
\left[\frac{b^2-\exp\left(\frac{\alpha
(t+e)}{bd}\right)}{r^2-\exp\left(\frac{\alpha (t+e)}{bd}\right)}
\right]^2 \left[ dr^2 + r^2 (d\theta^2 + \sin^2\theta d\phi^2)
\right]
\end{equation}
in terms of exponential functions. We believe that this is a new
solution to the Einstein field equations for a radiating star. It is
interesting to observe that if we set $\alpha =d$ when $t\rightarrow
\infty$ (or large values of the constant $e$) then (\ref{eq:e15})
becomes the flat Minkowski spacetime
\[ds^2 = - dt^2 +
dr^2 + r^2 (d\theta^2 + \sin^2\theta d\phi^2) \] which is a limiting
case.

When $u=\beta (t+\gamma)^{-2/3}$,  (\ref{eq:e13}) becomes
\begin{equation}
\label{eq:e16}
 \dot{C_1} +\left[\frac{\beta}{bd}
 (t+\gamma)^{-2/3}+\frac{2}{3}(t+\gamma)^{-1}\right]C_1 -\frac{b}{d}C_1^2=0
\end{equation}
which is also a Bernoulli equation with variable coefficients. The
solution of (\ref{eq:e16}) is given by
\[ C_1=\frac{\beta}{\left[b^2+\beta f
\exp \left(\frac{3 \beta
(t+\gamma)^{1/3}}{bd}\right)\right]}(t+\gamma)^{-2/3}\] where $f$ is
the constant of integration. Consequently the remaining function
$C_2$ is given by
\[C_2=\frac{-{\beta}^2f\exp \left(\frac{3 \beta
(t+\gamma)^{1/3}}{bd}\right) }{\left[b^2+\beta f \exp \left(\frac{3
\beta (t+\gamma)^{1/3}}{bd}\right)\right]}(t+\gamma)^{-2/3}.\] Hence
the interior line element (\ref{eq:e1}) takes the particular form
\begin{equation}
\label{eq:e17} ds^2 = - dt^2 + \frac{d^2}{{\beta}^2}
\left[\frac{b^2+\beta f \exp \left(\frac{3 \beta
(t+\gamma)^{1/3}}{bd}\right)} {r^2+\beta f \exp \left(\frac{3 \beta
(t+\gamma)^{1/3}}{bd}\right)}\right]^2(t+\gamma)^{4/3} \left[ dr^2 +
r^2 (d\theta^2 + \sin^2\theta d\phi^2)\right].
\end{equation}
If we set
\[\gamma=0, d=\left(\frac{M}{6}\right)^{1/3}b,
f= \frac{3}{a b^2}, \beta =-\frac{b^2}{3}\]
then (\ref{eq:e17}) becomes
\[ds^2 = - dt^2 + \frac{9 \left(\frac{M}{6}\right)^{2/3}}{b^2}
\left[\frac{1-ab^2 \exp \left(\frac{6t}{M}\right)^{1/3}}{1-ar^2 \exp
\left(\frac{6t}{M}\right)^{1/3}}\right]^2t^{4/3} \left[ dr^2 + r^2
(d\theta^2 + \sin^2\theta d\phi^2)\right]
\]
which was first found by  Kolassis \emph{et al} [11]. Here we have
shown that their model found by inspection  arises naturally as a
solution of a Bernoulli equation. It is easy to see that for large
values of the constant $f$ we obtain
\[ds^2 = - dt^2 + t^{4/3} \left[
dr^2 + r^2 (d\theta^2 + \sin^2\theta d\phi^2)\right]
\] from (\ref{eq:e17}). This corresponds to  the Friedmann metric when the
fluid is in the form of dust with vanishing heat flux.

\subsection{Riccati equation}
If we set
\begin{equation}
\label{eq:e18}u^2-db \dot{u}=0
\end{equation}
then the function $u$ is given by
\begin{equation}
\label{eq:e19} u =-bd(t+a)^{-1}
\end{equation}
where $a$ is a constant. In this case equation (\ref{eq:e10})
becomes
\begin{equation}
\label{eq:e20} 4bd \dot{C_1}-4b^2C_1^2+d^2(t+a)^{-2}=0,
\end{equation}
which is an inhomogeneous Riccati equation.  The solution of
equation (\ref{eq:e20}) has the form
\begin{equation}
\label{eq:e21} C_1= \frac{-d \left[1-\sqrt{2}+ (1+\sqrt{2})g
(t+a)^{\sqrt{2}}\right]}{2b \left[1+g
(t+a)^{\sqrt{2}}\right]}(t+a)^{-1},
\end{equation}
where $g$ is the constant of integration. Consequently the remaining
function has the form
\[C_2=bd\left\{1-\frac{ \left[1-\sqrt{2}+ (1+\sqrt{2})g
(t+a)^{\sqrt{2}}\right]}{2 \left[1+g
(t+a)^{\sqrt{2}}\right]}\right\}(t+a)^{-1}.\] Hence the interior
metric (\ref{eq:e1}) has the specific form
\begin{equation}
\label{eq:e22} ds^2 = - dt^2 + \frac{d^2 (t+a)^2}{\left[C_1 (r^2
-b^2)(t+a)-bd\right]^2} \left[ dr^2 + r^2 (d\theta^2 + \sin^2\theta
d\phi^2)\right],
\end{equation}
which is written in terms of $C_1$. We believe that (\ref{eq:e22})
is a new solution for a radiating star whose particles are
constrained to travel on geodesics. The simple form of
(\ref{eq:e22}) will assist in studying the physical features of our
model. The solution (\ref{eq:e22}) arises in a natural way once we
realise that the underlying dynamical equation (\ref{eq:e8}) at the
boundary is a Riccati equation.

\section{Special functions}
The  solutions found in the previous sections all have power law
forms for the quantity $u$. It is possible that other solutions in
terms of elementary functions or special functions may exist with a
power law representation for $u$. Consequently in this section we
attempt to generate a general class of solutions to the model
(\ref{eq:e8}) by assuming
\begin{equation}
\label{eq:e23} u =\alpha (t+a)^n.
\end{equation}
On substituting (\ref{eq:e23}) into (\ref{eq:e10}) we obtain
\begin{equation}
\label{eq:e24} (t+a)^2 \dot{C_1}+ \left[\frac{\alpha}{db}
(t+a)^{n+1} -n \right] (t+a)C_1 - \frac{b}{d}(t+a)^2 C_1^2=-
\frac{d}{4 b} n (3n+2).
\end{equation}
The nonlinear equation (\ref{eq:e24}) is a Riccati equation and it
is difficult to solve the equation  in the above  form. If we
introduce a transformation
\begin{equation}
\label{eq:e25} \frac{b}{d}C_1=-\frac{\dot{U}}{U}
\end{equation}
then (\ref{eq:e24}) becomes the second order linear differential
equation
\begin{equation}
\label{eq:e26} (t+a)^2 \ddot{U}+\left[\frac{\alpha}{db} (t+a)^{n+1}
-n \right](t+a) \dot{U}-\frac{n(3n+2)}{4}U=0
\end{equation}
in the function $U$ with variable coefficients. We can transform
(\ref{eq:e26}) to simpler form if we let
\begin{equation}
\label{eq:e27} \psi= (t+a)^{n+1}, ~ W=U \psi^{-k}, ~
k=\frac{(n+1)\pm\sqrt{4n(n+1)+1}}{2(n+1)}.
\end{equation}
Then (\ref{eq:e26}) becomes
\begin{equation}
\label{eq:e28} (n+1) \psi \frac{d^2W}{d {\psi}^2}
+\left[\frac{\alpha}{b d}\psi +2 k (n+1)\right]\frac{dW}{d\psi}
+\frac{\alpha k}{bd}W=0.
\end{equation}
If we let \[X=\frac{-\alpha\psi}{bd(n+1)},~Y(X)=W(\psi)\] then
(\ref{eq:e28}) has the equivalent form
\begin{equation}
\label{eq:e29} X\frac{d^2Y}{dX^2} +(2k-X)\frac{dY}{dX}-kY=0.
\end{equation}
Observe that (\ref{eq:e29}) is the confluent hypergeometric equation
with solution in terms of special functions in general.

Note that the solution of (\ref{eq:e29}) can be written in terms of
\begin{eqnarray}
Y&=&\mathcal{J}\left(k,2k;X \right),\nonumber\\
W&=&\mathcal{J}\left(k,2k;\frac{-\alpha\psi}{bd(n+1)}\right)\nonumber
\end{eqnarray}
where $\mathcal{J}$  are Kummer functions. In general the solution
of the equation (\ref{eq:e28}) can be written in terms of the Kummer
series. Observe that when $k>0$ we can write
\begin{eqnarray}
 \tilde{W} &=& \mathcal{J}(k,2k;X)\nonumber\\
\label{eq:e30}&=&\frac{\Gamma(2k)}{\left[\Gamma(k)\right]^2}\int_0^1
e^{X\tau} \left[\tau (1-\tau)\right]^{k-1}d\tau
\end{eqnarray}
as a  particular solution of the differential equation
(\ref{eq:e28}) where $\Gamma(z)=\int_0^\infty e^{-\tau}\tau^{z-1}
d\tau$ is the gamma function.  From (\ref{eq:e30}) we note that the
solution can be expressed in terms of elementary functions for all
natural numbers $k$. Consequently the  differential equation
(\ref{eq:e24}) admits
solutions in terms of elementary functions when $k$ is a natural number.\\

\subsection{Particular metrics}
We can regain previous cases from the general form (\ref{eq:e30}).
We illustrate this feature for particular values of $k$. When $k=1$,
we obtain $n=0$ or $n=-2/3$. For this case the particular solution
of the equation (\ref{eq:e28}) becomes
\begin{equation}
\label{eq:e31}\tilde{ W }= \frac{e^X -1}{X},~~~X=\frac{-
\alpha\psi}{b d (n+1)}
\end{equation}
with the help of (\ref{eq:e30}).\\

When $n=0$, from (\ref{eq:e27}) and (\ref{eq:e31}) we can easily see
that
\begin{equation}
\tilde{U}= \frac{bd}{\alpha}\left[1-\exp\left({\frac{-\alpha
(t+a)}{b d}}\right)\right]\nonumber
\end{equation}
is a particular solution of the equation (\ref{eq:e26}). Then with
the help of (\ref{eq:e25}) we find that
\begin{equation}
\label{eq:e32}\tilde{C_1}= \frac{\alpha}{b^2 \left[1-
\exp\left({\frac{\alpha (t+a)}{b d}}\right)\right]}
\end{equation}
is a particular solution of (\ref{eq:e24}) which is given by
\begin{equation}
\label{eq:e33}\dot{C_1}+\frac{\alpha}{bd}C_1 -\frac{b}{d}{C_1}^2=0.
\end{equation}
The general solution of (\ref{eq:e33}) becomes
 \[C_1 =\frac{\alpha D}{b^2\left[ D+
 \exp{\left(\frac{\alpha (t+a)}{bd}\right)}\right]}, \] where $D$ is
 an arbitrary constant.  Consequently the interior metric (\ref{eq:e1}) has the specific form
\begin{equation}
\label{eq:e34} ds^2 = - dt^2 + \frac{ b^4d^2}{{\alpha}^2}
\left[\frac{D + \exp{\left(\frac{\alpha (t+a)}{bd}\right)}} {Dr^2
+b^2 \exp{\left(\frac{\alpha (t+a)}{bd}\right)}}\right]^2
 \left[ dr^2 + r^2 (d\theta^2 + \sin^2\theta d\phi^2)\right]
\end{equation}
in terms of exponential functions. Note that the line element
(\ref{eq:e34}) reduces to the metric (\ref{eq:e15}) if we set
$D=-b^2$.

When $n=-2/3$, from (\ref{eq:e27}) and (\ref{eq:e31}) we observe
that
\begin{equation}
\tilde{U}= \frac{bd}{3 \alpha}\left[1-\exp\left({\frac{-3 \alpha
(t+a)^{1/3}}{b d}}\right)\right]\nonumber
\end{equation}
is a particular solution of the equation (\ref{eq:e26}). Hence with
the help of (\ref{eq:e25}) we obtain
\begin{equation}
\label{eq:e35}\tilde{C_1}= \frac{\alpha (t+a)^{2/3}}{b^2 \left[1-
\exp\left({\frac{3 \alpha (t+a)^{1/3}}{b d}}\right)\right]}
\end{equation}
as a particular solution of (\ref{eq:e24}) which has the form
\begin{equation}
\label{eq:e36}\dot{C_1}+ \left[\frac{\alpha}{bd} (t+a)^{-2/3}+
\frac{2}{3}(t+a)^{-1}\right]C_1-\frac{b}{d}{C_1}^2=0.
\end{equation}
The general solution of (\ref{eq:e36}) becomes
 \[C_1= \frac{\alpha D(t+a)^{2/3}}{b^2 \left[ D +
 \exp{\left(\frac{3\alpha (t+a)^{1/3}}{bd}\right)}\right]}, \]
where $D$ is an arbitrary constant. Consequently the  interior
metric (\ref{eq:e1}) takes the particular form
\begin{equation}
\label{eq:e37} ds^2 = - dt^2 +  \frac{b^4d^2}{{\alpha}^2}
\left[\frac{D +
 \exp{\left(\frac{3\alpha (t+a)^{1/3}}{bd}\right)}}{D r^2 +
 \exp{\left(\frac{3\alpha (t+a)^{1/3}}{bd}\right)}}\right]^2 (t+a)^{4/3} \left[ dr^2 +
r^2 (d\theta^2 + \sin^2\theta d\phi^2)\right].
\end{equation}
Note that the line element (\ref{eq:e37}) reduces to the line
element (\ref{eq:e17}) if we set $D=\frac{b^2}{\alpha f}$.

\subsection{A new solution}
It is possible to generate an infinite family of new solutions from
the general form (\ref{eq:e30}) by specifying values for the
parameter $k$. These may  correspond to new solutions for a
radiating sphere which are not accelerating. We illustrate this
process by taking $k=2$ (so that $n=-2$ or $n=-4/5$) in
(\ref{eq:e30}). We consider only the case $n=-2$ as the integration
procedure is same for other values of $k~(\mbox{or}~n)$ . For this
case the particular solution of the equation (\ref{eq:e28}) becomes
\begin{equation}
\label{eq:e38} \tilde{W} =\frac{6}{X^3} \left[2+ X + (X-2)e^X
\right],~~~X=\frac{- \alpha\psi}{b d (n+1)}.
\end{equation}
When $n=-2$, from (\ref{eq:e27}) and (\ref{eq:e38}) we observe that
\begin{equation}
\tilde{U}= \frac{6 b^2 d^2}{\alpha^3} \left[[2bd (t+a)+\alpha] -[2bd
(t+a)-\alpha]\exp\left({\frac{\alpha}{bd
(t+a)}}\right)\right]\nonumber
\end{equation}
is a particular solution of the equation (\ref{eq:e26}). Hence with
the help of (\ref{eq:e25}) we obtain
\begin{equation}
\label{eq:e39}\tilde{C_1}= \frac{2 b^2 d^2 (t+a)^2 -\left[2b d
(t+a)(bd (t+a)-\alpha) +\alpha^2\right]\exp{\left(\frac{\alpha}{bd
(t+a)}\right)}} {b^2 \left[(2bd
(t+a)-\alpha)\exp{\left(\frac{\alpha}{bd (t+a)}\right)}-(2bd
(t+a)+\alpha)\right](t+a)^2}
\end{equation}
is a particular solution of (\ref{eq:e24}) which has the form
\begin{equation}
\label{eq:e40} \dot{C_1}+ \left[\frac{\alpha}{bd}(t+a)^{-2}+ 2
(t+a)^{-1}\right]C_1 -\frac{b}{d}{C_1}^2 =-\frac{2d}{b}(t+a)^{-2}.
\end{equation}
The general solution of (\ref{eq:e40}) becomes
\begin{equation}
\label{eq:e41}C_1 =- \frac{\left[ \left[\alpha^2 +2bd (t+a)
(bd(t+a)-\alpha)\right]\exp{\left(\frac{\alpha}{bd(t+a)}\right)}+2b^2
d^2D (t+a)^2 \right]}{b^2 \left[D(2bd(t+a)+\alpha)+
(2bd(t+a)-\alpha)\exp{\left(\frac{\alpha}{bd(t+a)}\right)}\right](t+a)^2},
\end{equation}
where $D$ is an arbitrary constant.  Consequently the  interior
metric (\ref{eq:e1}) has the specific form
\begin{equation}
\label{eq:e42} ds^2 = - dt^2 + \frac{d^2}{\left[C_1(r^2 -b^2)+\alpha
(t+a)^{-2}\right]^2} \left[ dr^2 + r^2 (d\theta^2 + \sin^2\theta
d\phi^2)\right],
\end{equation}
where $C_1$ is given by (\ref{eq:e41}). Hence we have found a new
solution to the boundary condition (\ref{eq:e8}) by specifying a
particular value for the parameter $k$. This process can be repeated
for other values of $k$ and an infinite family of solutions are
possible in which the gravitational potentials can be expressed in
terms of elementary functions.

\section{Physical analysis}
The simple forms of the gravitational potentials found in this paper
permit a detailed study of the physical features of a radiating
star. In this study we consider the particular line element
(\ref{eq:e34}) and set $\alpha = b d$ and $a=0$ to obtain
\begin{equation}
\label{eq:e43}ds^2= - dt^2 + b^2 \left[\frac{D + \exp(t)} {Dr^2 +b^2
\exp(t)}\right]^2
 \left[ dr^2 + r^2 (d\theta^2 + \sin^2\theta d\phi^2)\right],
\end{equation}
for simplicity. For the metric (\ref{eq:e43}) the matter variables
can be written as
\begin{subequations}
\label{eq:e44}
\begin{eqnarray}
\label{eq:e44a}\rho  &=&\frac{3 D  \exp (t)\left\{\exp (t)
\left[Db^4+6b^2Dr^2+Dr^4 +4b^4 \exp (t)\right]+4D^2r^4\right\}}{
\left[D+\exp (t)\right]^2 \left[Dr^2+ b^2
\exp (t)\right]^2},\\
 p &=& \frac{D \exp (t)}{ \left[D+\exp (t)\right]^2 \left[Dr^2+ b^2
\exp (t)\right]^2}\times \nonumber\\
&&\left\{\exp (t))\left[2b^2\exp (t)(r^2-3b^2)-2Db^2r^2-
3D(r^4+b^4)\right] +2D^2r^2(b^2-3r^2)\right\},\nonumber\\
\label{eq:e44b}&&\\
\label{eq:e44c} q&=&\frac{4Dr\exp (t)}{\left[D+\exp (t)\right]^2}.
\end{eqnarray}
\end{subequations}
When $D=0$ then (\ref{eq:e43}) becomes the Minkowski metric with
$\rho =p=q=0$.  The matter variables are expressed in simple
analytic forms which facilitate the analysis of the physical
behaviour. From (\ref{eq:e44}) we have that at the centre of the
sphere
\begin{eqnarray}
{\rho}_0 &=&\frac{3 D  \left[D +4 \exp (t)\right]}{ \left[D+\exp
(t)\right]^2 } \nonumber\\
p_0 &=& -\frac{3D[D+2 \exp (t)]}{[D+\exp (t)]^2}\nonumber\\
q_0 &=&0 \nonumber
\end{eqnarray}
so that ${\rho}_0$ and $p_0$ have finite values at the centre $r=0$
with vanishing heat flux $q_0 $.  The gravitational potentials in
(\ref{eq:e43}) are finite at the centre and nonsingular in the
stellar interior. The quantities $\rho, p$ and $q$ are well behaved
and regular in the interior of the sphere, at least in regions close
to the centre. At later times as $t \rightarrow \infty$ we note that
$q \propto r$ so that the magnitude of the heat flux depends
linearly on the radial coordinate.

Next we briefly consider the relativistic effect of casual
temperature of this model. The Maxwell-Cattaneo heat transport
equation, in the absence of rotation and viscous stresses, is given
by
\begin{equation}
\label{eq:e45} \tau h_a^{~b}\dot{q}_b+q_a = -\kappa
\left(h_a^{~b}{\nabla}_b T+T \dot{u}_a\right),
\end{equation}
where $h_{ab}=g_{ab}+u_au_b$ projects into the comoving rest space,
$T$ is the local equilibrium temperature, $\kappa ~(\geq 0)$ is the
thermal conductivity and $\tau ~(\geq 0)$ is the relaxation  time.
Equation (\ref{eq:e45}) reduces to the acausal Fourier heat
transport equation when $\tau =0$. For the line element
(\ref{eq:e1}), the casual transport equation (\ref{eq:e45}) can be
written as
\begin{equation}
\label{eq:e46} T(t,r)= -\frac{1}{\kappa } \int \left[\tau
\dot{(qB)}B+ qB^2 \right]dr
\end{equation}
for geodesic motion. Martinez [19], Govender \emph{et al} [16] and
Di Prisco \emph{et al} [20] have demonstrated that the relaxation
time $\tau$ on the thermal evolution, plays a significant role in
the latter stages of collapse. For the line element (\ref{eq:e43}),
(\ref{eq:e46}) becomes
\begin{eqnarray}
\label{eq:e47}T(t,r)&=&\frac{\tau b^2
\exp(t)\left\{2D^2r^2-b^2\exp(t)\left[\exp(t)-D\right]\right\}}{\kappa
\left[\exp(t) +D\right]\left[b^2\exp(t)+Dr^2\right]^2} \nonumber\\
&& +\frac{2  b^2 \exp(t)}{\kappa \left[b^2\exp(t)+Dr^2\right]}+h(t),
\end{eqnarray}
where $h(t)$ is a function of integration. For simplicity we assumed
that $\tau$ and $\kappa$ are constant. The function $h(t)$ may be
related to the central temperature $T_c(t)$ by
\begin{equation}
\label{eq:e48} h(t)= T_c(t)- \frac{\tau
\left[D-\exp(t)\right]}{\kappa \left[D+\exp(t)\right]}-\frac{2
}{\kappa }.
\end{equation}
From (\ref{eq:e47}) and (\ref{eq:e48})  the temperature can be
written as
\begin{eqnarray}
\label{eq:e49} T(t,r)&=& T_c(t)-\frac{\tau  D r^2 \left\{Dr^2
\left[D-\exp(t)\right]-2b^2\exp(t)\right\}}{\kappa
\left[D+\exp(t)\right]\left[Dr^2+b^2\exp(t)\right]^2}\nonumber \\
&& - \frac{2  Dr^2}{\kappa  \left[Dr^2+b^2 \exp(t)\right]}.
\end{eqnarray}
When $\tau =0$, we can regain the acausal (Eckart) temperature
profiles from (\ref{eq:e49}). In Fig. 1, we plot the casual (solid
line) and acasual (dashed line) temperatures against the radial
coordinate on the interval $0 \leq r \leq 5$ for particular
parameter values $(\kappa = \tau =1,  b =5, D=70 ~\mbox{and}~
h(t)=0)$ on the spacelike hypersurface $t=1$ . We  observe that the
temperature is monotonically decreasing from centre to the boundary
in both casual and acasual cases. It is clear that the casual
temperature is greater than the acasual temperature throughout the
stellar interior. At the boundary $\Sigma$ we have
\[ T(t,r_{\Sigma})_{\mbox{casual}} \simeq
T(t,r_{\Sigma})_{\mbox{acasual}}.\] Our figures have been generated
by assuming constant values for the parameters $\tau $ and $\kappa$.
Changing the values of the relaxation time and the thermal
conductivity would produce different gradients for the curves but
the result would not change qualitatively.

\vspace{1cm}
\begin{figure}[thb]
\vspace{1.5in} \includegraphics{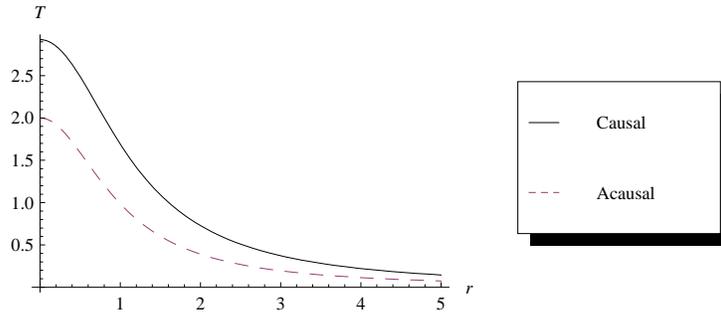}
\caption{\label{Temp-graph-23-new} Temperature $T$ vs radial
coordinate $r~(\tau =1)$.}
\end{figure}

\section{Discussion}
It is possible to introduce shear in geodesic motion as shown by
Naidu \emph{et al} [17] and Rajah and Maharaj [18] in the
description of a radiating star. The solutions that follow are
governed by a Riccati equation and have a complicated form.
Consequently in this paper we have considered the simpler case of a
shear-free metric with particles traveling on geodesic trajectories.
The master equation, governing the boundary condition of the stellar
model, was transformed to a simpler form. Under certain assumptions
a Bernoulli equation is possible. This Bernoulli equation admits two
solutions in terms of elementary functions: the first solution
contains the Minkowski spacetime as a limiting case and the second
solution corresponds to the Kolassis \emph{et al} [11] model with
the Friedmann dust spacetime as  the limiting case. A general class
of solutions are possible if we transform the master equation to a
confluent hypergeometric equation. The resulting transformed
equation admits solution in terms of special functions namely the
Kummer functions. By specifying particular values for a parameter in
the special function we demonstrate that an infinite family of
solutions, in terms of elementary functions, are possible. The
simple form of the solutions makes it possible to study the physical
features of the model and to find an analytic form for the causal
temperature.

\section*{Acknowledgements}
ST thanks the National Research Foundation and the University of
KwaZulu-Natal for financial support, and is grateful to Eastern
University, Sri Lanka for study leave. SDM acknowledges that this
work is based upon research supported by the South African Research
Chair Initiative of the Department of Science and Technology and the
National Research Foundation.

\end{document}